\documentclass[apsrev,superscriptaddress,showpacs]{revtex4}
\begin{document}
\title{ Non-Standard Probabilistic Teleportation through Conventionally Non-Teleporting Channels }
\author{Mayank Mishra}
\affiliation{ Indian Institute of Science Education and Research Mohali, Transit Campus: MGSIPAP Complex, Sector 26 Chandigarh-160019, India}
\author{Atul Mantri}
\affiliation{ Indian Institute of Science Education and Research Mohali, Transit Campus: MGSIPAP Complex, Sector 26 Chandigarh-160019, India}
\author{Priyank Mishra}
\affiliation{National Institute of Engineering, Mysore-570008, India }
\email{priyank1119@gmail.com}
\author{P.K. Panigrahi}
\affiliation{Indian Institute of Science Education and Research Kolkata, Mohanpur Campus, Nadia-741252, West Bengal, India}
\email{prasanta@prl.res.in}
\begin{abstract}
A non-standard teleportation scheme is proposed, wherein probabilistic teleportation is achieved in conventionally non-teleporting channels. We make use of entanglement monogamy to incorporate an unknown state in a multipartite entangled channel, such that the receiver partially gets disentangled from the network. Subsequently, the sender performs local measurement based teleportation protocol in an appropriate measurement basis, which results with the receiver in the possession of an unknown state, connected by local unitary transformation with the state to be teleported. This procedure succeeds in a number of cases, like that of W and other non-maximally entangled four qubit states, where the conventional measurement based approach has failed. It is also found that in certain four particle channels, the present procedure does not succeed, although the conventional one works well.
\end{abstract} 
\maketitle

\section{Introduction}
One of the intriguing features of quantum mechanics is entanglement, which has been exploited to carry out a number of computational and information processing tasks, which are otherwise impossible. These include teleportation of an unknown quantum state \cite{1}, superdense coding \cite{2}, entanglement swapping \cite{3}, secret sharing \cite{4}, quantum cryptography \cite{5}, secure quantum conversation \cite{6}, non-destructive discrimination of entangled state \cite{7} and many other operations \cite{8}. Teleporation is a spectacular application of entanglement, where information about an unknown state can be teleported, with complete fidelity, to a distant receiver in an entangled channel. In the celebrated teleportation protocol of Bennett et al. \cite{1}, the sender Alice performs a joint Bell measurement, involving the unknown qubit and her own state. Subsequent to the measurement, Alice's measured state decouples from the receiver Bob, leaving his state in an unknown
  superposed state, connected by a local unitary transformation with the desired state. Suitable classical communication from Alice, depending on the experimentally projected states, enables Bob to retrieve the exact state, after applying the required unitary transformation. This measurement based teleportation protocol (MBTP) has found experimental verification \cite{9,10} and has been extended to a number of multi-partite entangled channels \cite{11,12,13,14,15,16}, for teleportation of single and multi-qubit unknown states. As is evident, the nature of entanglement plays a crucial role in MBTP. Whereas the maximally entangled Bell state can teleport with certainty using LOCC, the corresponding non-maximally entangled channel can achieve the same probabilistically. It is known that, if the shared resource is a non-maximally entangled state then one has to follow a scheme, in which teleportation  of a qubit is possible with unit fidelity and non-unit probability \cite{17}. In case of three particles, it has been found that the three particle GHZ state can teleport successfully, whereas the W-state fails completely \cite{18} in the sense that, the measurement basis at the sender's end necessarily depends on the state to be teleportated. In general, the three particle state can be categorized into two classes from the consideration of teleportation \cite{19}, whereas the four particle states form nine classes under SLOCC \cite{6,20,21,27}.\\ It is worth mentioning that a number of schemes have been designed to make W and related states useful for teleportation \cite{22,23,24}. In the proposal of Ref. \cite{22}, teleportation can be successfully realized with a certain probability, if the receiver adopts an appropriate unitary-reduction strategy, without using local operation. In the scheme of Ref. \cite{23}, a sender performs positive operator valued measurement (POVM) to realize teleportation of the single particle state, when the channel is shared between three parties. After the first party measures his state, the channel breaks up into a two particle state, which can be used for probabilistic teleportation. In Ref. \cite{24} probabilistic teleportation of an unknown two particle entangled state has been carried out through W-state as the entangled channel, wherein the receiver measures one of his particle before the unitary transformation and then adds an ancilla to get the desired state through local operation and POVM. Teleportation for the purpose of quantum information splitting in four and higher particle particle states have been investigated \cite{25}, where a number of four particle entangled states have been found unsuitable for the same.\\
In this paper, we present a non-standard teleportation scheme, in which probabilistic teleportation is achieved in several conventionally non-teleporting channels, like that of W and other non-maximally entangled four qubit states, where the conventional measurement based approach has failed. It is also found that, in certain four particle channels the present procedure succeeds probabilistically, although the conventional approach works deterministically. We make use of the monogamy property of the entangled channel\cite{28}, because of which the incorporation of an unknown state into a quantum communication network, partially decouples the sender from the network, leaving the receiver in a state unitarily equivalent to the unknown one. Subsequent measurement at Alice's end, involving either computational basis or entangled basis yields the desired information to be classically communicated to Bob for implementation of the necessary local operations. \\ The paper is organized as follows. In the following section, we explicate the present approach through the Bell state and then proceed to the three particle W-state, to show probabilistic teleportation. Sec. III is devoted to the study of teleportation  in a number of four particle states, earlier found unsuitable for teleportation through conventional MBTP approach. We conclude in Sec. IV, after pointing out a number of directions for future work.

\section{PROBABILISTIC TELEPORTATION THROUGH W-STATE}
 We start with the illustration of the present approach, using Bell state $|\Psi^{+}\rangle_{AB} = \frac {1} {\sqrt{2}}(|00\rangle + |11\rangle)_{AB}  $ as a quantum channel, between Alice and Bob. For this purpose any type of Bell state can be used. At the Alice end, one can incorporate the unknown state $\alpha|0\rangle +\beta|1\rangle$, ( $|\alpha|^{2} + |\beta|^{2} = 1 $) to the entangled channel, through the application of a control-NOT operation, with unknown state as control qubit and the first particle of $|\Psi^{+}\rangle_{AB}$ as target. Alice then applies a Hadamard operation on the first qubit. In this process, Alice is projected onto computational basis, leaving Bob with the possession of an unknown state, unitarily connected with the state to be teleported. Subsequent uncorrelated von Neumann measurement by Alice and communication of the same in a classical channel through two c-bits, enables Bob to apply local unitary transformations on his state to recover the
  desired state. For the Bell state, this present protocol and the conventional measurement based approach \cite{8} yield  identical result. However, as will be explicitly seen below, the same is not true for multi-partite cases. We now proceed to study the utility of W-state for teleportation using the present scheme.

\subsection{Probabilistic Teleportation of single qubit unknown state through $|W \rangle $ state }
For the $|W \rangle$ state,
\begin{equation}
|W \rangle_{234} =  \frac {1} {\sqrt{3}} [ | 100 \rangle_{234}  + | 010 \rangle_{234}  + |001 \rangle_{234}],
\end{equation} 
the particles 2 and 3 belong to Alice and the particle 4 belongs to Bob. With addition of the unknown state at Alice's end, the quantum channel takes the form, 
\begin{eqnarray}
|\Psi \rangle_1 \otimes |W \rangle_{234} =\frac {1} {\sqrt{3}}  [\alpha |0100 \rangle_{1234} + \alpha |0010 \rangle_{1234} + \alpha |0001 \rangle_{1234} + \beta |1100 \rangle_{1234} + \beta |1010 \rangle_{1234} + \beta |1001\rangle_{1234} ]. 
\end{eqnarray}
 For the incorporation of the unknown state into the entangled channel, Alice applies a control-NOT, with particle 1 as control qubit and particle 2 as the target. Subsequently, a Hadamard gate on the first qubit is applied by Alice, which leads to,
\begin{eqnarray}
|\Psi \prime\prime\rangle_{1234} = \frac {1} {\sqrt{6}} [ |010 \rangle_{123} ( \alpha|0 \rangle_4 + \beta |1\rangle_4) + |000 \rangle_{123}(\alpha |1 \rangle_4 + \beta |0 \rangle_4) + |110 \rangle_{123}( \alpha|0 \rangle_4  -\beta |1 \rangle_4) \nonumber\\+ |100 \rangle_{123}( \alpha|1 \rangle_4 - \beta  |0 \rangle_4)  + \alpha |001 \rangle_{123} |0 \rangle_4 + \alpha |101 \rangle_{123} |0 \rangle_4 + \beta |011 \rangle_{123} |0 \rangle_4 - \beta |111 \rangle_{123} |0 \rangle_4 ].
\end{eqnarray}
 Alice then performs a three particle measurement on her qubits in the computational basis and communicates the obtained result to Bob, so that he carries out the required operations on his qubit to get the desired state. It is evident that teleportation is possible, only if the measurement outcomes are $|010 \rangle_{123}$, $|000 \rangle_{123}$, $|110 \rangle_{123}$ and $|100 \rangle_{123}$ and fails completely if the outcomes are $|001 \rangle_{123}$, $|101 \rangle_{123}$, $|011 \rangle_{123}$ and $|111 \rangle_{123}$. Hence probability of teleportation is half. It can be check that for the GHZ state this protocol works deterministically.
\subsection{Teleportation of two qubit non-maximally entangled unknown state through $|W \rangle$ channel }
	We now proceed for the probabilistic teleportation of two qubits through W-state. It has been known that probabilistic teleportation of an unknown two qubit entangled state can be carried out through W-state as the entanged channel, if  Bob measures one of his particle before the required unitary transformation \cite{24}. In the present scheme, teleportation occurs without the interaction of Alice's qubits with Bob's qubits. In present procedure, general unknown two-qubit state cannot be teleported using W-state as a quantum resource \cite{25}, however probabilistic teleportation of non-maximally entangled two qubit state of the type: $|\Psi \rangle_{12} = \alpha |00 \rangle +\beta [|01 \rangle+ |10 \rangle]$ is possible ( $|\alpha|^{2} + 2|\beta|^{2} = 1$).
We assume that particle 3  belongs to Alice and the particles 4,5 belong to Bob:
\begin{eqnarray}
|\Psi \rangle_{12} \otimes |W \rangle_{345} =\frac {1} {\sqrt{3}} [\alpha |00001 \rangle_{12345} + \alpha |00010 \rangle_{12345} + \alpha |00100\rangle_{12345} + \beta |10001 \rangle_{12345} + \nonumber \\\beta |10010 \rangle_{12345} + \beta |10100\rangle_{12345} +\beta |01001\rangle_{12345}+\beta |01010\rangle_{12345} + \beta |01100 \rangle_{12345}].  
\end{eqnarray}
Now Alice applies a control-NOT on 3  as target qubit and 2 as control qubit and subsequently another control-NOT is applied with 3 as target qubit and 1 as control qubit. After that Alice carries out a measurement on the qubit 3 in computational basis:
\begin{eqnarray}
\langle1_3|\Psi \prime\rangle_{12345} =\frac {1} {\sqrt{|\alpha|^2 + 4|\beta|^2}} [\alpha |0000\rangle_{1245}+\beta |1010 \rangle_{1245}+\beta |0100\rangle_{1245}+\beta |0110\rangle_{1245}+\beta |1001\rangle_{1245}],
\end{eqnarray}
or
\begin{eqnarray}
\langle0_3|\Psi \prime\rangle_{12345} =\frac {1} {\sqrt{2|\alpha|^2 + 2|\beta|^2}} [\alpha |0001 \rangle_{1245} + \alpha |0010 \rangle_{1245} + \beta |1000 \rangle_{1245} +\beta |0100 \rangle_{1245}].
\end{eqnarray}
Teleportation is possible if the measurement outcome is $|1_{3}\rangle$ and fails if it is $|0_{3}\rangle$. In the former case, Alice applies a Hadamard on qubit 1 and then measures qubit 2 in computational basis, leading to two outcomes:
\begin{eqnarray}
\langle0_2|\Psi \prime\prime\rangle_{1245} =\frac {1} {\sqrt{2}} [(|0\rangle_{1})(\beta |01\rangle_{45}+\beta|10\rangle_{45})-(|1\rangle_{1})(\beta |01\rangle_{45}+\beta|10\rangle_{45})\nonumber\\+(|0\rangle_{1})(\alpha |00\rangle_{45})+(|1\rangle_{1})\alpha|00\rangle_{45})],
\end{eqnarray}
or
\begin{eqnarray}
\langle1_2|\Psi \prime\prime\rangle_{1245} =\frac {1} {2|\beta|} [(|0\rangle_{1})(\beta |00\rangle_{45}+\beta|10\rangle_{45})+(|1\rangle_{1})(\beta |00\rangle_{45}+\beta|10\rangle_{45})].
\end{eqnarray}
It is clear that, teleportation is possible if the measurement outcome is $|0_{2}\rangle$ and fails if it is $|1_{2}\rangle$. In case of the first outcome, Alice measures qubit 1 in computational basis and sends the results of  her measurement outcome to Bob, so that he can make the required local operations to get the desired state. In the table 1, outcomes of the final measurement of Alice in computational basis are shown.\\
\begin{table}[h]
\caption{\label{tab1} The outcome of the measurement performed by Alice and the state obtained by Bob}
\begin{center}
\begin{tabular}{|c|c|}
\hline
{\bf Outcome of Alice's Measurement}&{\bf State obtained}\\
\hline
$|0\rangle$&$\alpha(|00\rangle_{45})+\beta(|01\rangle_{45}+|10\rangle_{45})$\\
$|1\rangle$&$\alpha(|00\rangle_{45})-\beta(|01\rangle_{45}+|10\rangle_{45})$\\
\hline
\end{tabular}
\end{center}
\end{table}
In table 1, if  the outcome of Alice's measurement is $|1\rangle$, then Bob applies a unitary transformation to obtain the desired  state.
Thus probabilistic teleportation of partially entangled two qubits have been implemented with probability of success 1/4. If one considers $|W \rangle_{234} =  \frac {1} {\sqrt{3}} [ | 101 \rangle_{234}  + | 110 \rangle_{234}  + |011 \rangle_{234} ]$ as entangled channel, then the state that can be teleportated is $|\Psi \rangle_{12} = \alpha |11 \rangle_{12} + \beta [|01 \rangle_{12} + |10 \rangle_{12}]$.\\
It is interesting to note that, when Alice fails to teleport the non-maximally entangled unknown two qubit state, she can regain this unknown two qubit state, with certain probability, by a classical communication with Bob about the failure of teleportation. Bob then applies a Hadamard operation on particle 5, followed by a measurement of his two particles in computational basis.  In table 2, outcomes of the measurement of Bob in computational basis are shown.\\
\begin{table}[h]
\caption{\label{tab2} The outcome of the measurement performed by Bob and the state obtained by Alice}
\begin{center}
\begin{tabular}{|c|c|}
\hline
{\bf Outcome of Bob's Measurement }&{\bf State obtained }\\
\hline
$|00\rangle_{45}$&$\alpha(|00\rangle_{12})+\beta(|01\rangle_{12}+|10\rangle_{12})$\\
$|01\rangle_{45}$&$\alpha(-|00\rangle_{12})+\beta(|01\rangle_{12}+|10\rangle_{12})$\\
$|11\rangle_{45}$&$|00\rangle_{12}$\\
$|10\rangle_{45}$&$|00\rangle_{12}$\\
\hline
\end{tabular}
\end{center}
\end{table} 
In table 2, if  the outcome of Bob's measurement is $|01\rangle$ or $|00\rangle$, then Alice applies required unitary transformations to obtain the desired  state with probability 1/2 and one-cbit of classical communication.\\

\section{Teleportation through quadripartite state }
On the basis of SLOCC classification, four particle states have been classified into different types \cite{20,21,27}. We consider a subset of states from this classification and investigate the possibility of implementing the present protocol of teleportation. These states are:
\begin{equation}
|P_{1} \rangle \equiv |W\rangle =  \frac {1} {2} [ | 0001 \rangle  + | 0010 \rangle  + |0100 \rangle + |1000 \rangle],
\end{equation}
\begin{equation}
|P_{2} \rangle =  \frac {1} {\sqrt {5}} [ | 0000 \rangle  + | 1111 \rangle  + |0011 \rangle + |0101 \rangle + |0110 \rangle],
\end{equation}
\begin{equation}
|P_{3} \rangle =  \frac {1} {2} [ | 0000 \rangle  + | 0101 \rangle  + |1000 \rangle + |1110 \rangle]
\end{equation}
and
\begin{equation}
|P_{4} \rangle =  \frac {1} {2} [ | 0000 \rangle  + | 1011 \rangle  + |1101 \rangle + |1110 \rangle].
\end{equation}
In the conventional MBTP approximation, teleportation of unkown one qubit state through $ |P_{1} \rangle$ and $|P_{2} \rangle$ and two qubit non-maximally entangled state through $|P_{1} \rangle$ and $|P_{2} \rangle$, has failed, whereas the present protocol results in a probabilistic teleportation. We note that $|P_{3}\rangle $ and $|P_{4}\rangle $ carry out faithful teleportation of single qubit, by standard protocol \cite{26}. It will be seen that, by applying local operations teleportation becomes probabilistic.
\subsubsection{Teleportation of single qubit unknown state through $|P_{1} \rangle $ state }
We start with,
\begin{equation}
|P_{1} \rangle_{3456} =  [\frac {1} {2} ][ | 0001 \rangle_{3456}  + | 0010 \rangle_{3456}  + |0100 \rangle_{3456} + |1000 \rangle_{3456}],
\end{equation} 
where the particles 3, 4 and 5 belong to Alice and the particle 6 belongs to Bob. With the addition of unknown state of one qubit at the Alice's end, the state can be written as,
\begin{eqnarray}
|\Psi \rangle_1 \otimes |P_{1} \rangle_{3456} =\frac {1} {2}  [\alpha |00001 \rangle_{13456} + \alpha |00010 \rangle_{13456} + \alpha |00100 \rangle_{13456} + \alpha |01000 \rangle_{13456} + \beta |10001 \rangle_{13456} \nonumber \\
  + \beta |10010\rangle_{13456} +\beta |10100\rangle_{13456}+\beta |11000\rangle_{13456}]. 
\end{eqnarray}
It is not possible to rewrite the above state in appropriate orthogonal measurement basis at the Alice's end, leaving it unsuitable for teleportation through standard protocol. Probabilistic teleportation can be done by using local operations. Alice applies control-NOT with 1 as a control qubit and 4 as a target qubit and then another control-NOT, with 2 as a control qubit and 3 as a target:
\begin{eqnarray}
|P_{1} \prime\rangle_{13456} =\frac {1} {2}  [\alpha |00001 \rangle_{13456} + \alpha |00010 \rangle_{13456} + \alpha |00100 \rangle_{13456} + \alpha |01100 \rangle_{13456} + \beta |10011 \rangle_{13456} + \nonumber \\\beta |10000\rangle_{13456} +\beta |10110\rangle_{13456}+\beta |11110\rangle_{13456}]. 
\end{eqnarray}
Now Alice measures 1, 2 and 3, 4 successively in Bell basis:
\begin{eqnarray}
\langle\Phi^{\pm}|=\frac {1}{\sqrt{2}}[\langle 00|\pm \langle11|]
\end{eqnarray}
and
\begin{eqnarray}
\langle\Psi^{\pm}|=\frac {1}{\sqrt{2}}[\langle 01|\pm\langle10|],
\end{eqnarray}
which respectively yields,
\begin{eqnarray}
\langle \Phi^{\pm}|_{45} \langle \Phi^{\pm}|_{13} P_{1} \prime\prime\rangle_{13456} =\alpha|1 \rangle_6 \pm \beta|0 \rangle_6,
\end{eqnarray}
\begin{eqnarray}
\langle \Psi^{\pm}|_{45}  \langle \Phi^{\pm}|_{13} P_{1} \prime\prime\rangle_{13456} = |0\rangle_6, 
\end{eqnarray}
\begin{eqnarray}
\langle \Phi^{\pm}|_{45}  \langle \Psi^{\pm}|_{13} P_{1} \prime\prime\rangle_{13456} = |0\rangle_6,
\end{eqnarray}
or
\begin{eqnarray}
\langle \Psi^{\pm}|_{45}  \langle \Psi^{\pm}|_{13} P_{1} \prime\prime\rangle_{13456} =\pm\alpha|0\rangle_6 \pm \beta|1 \rangle_6.
\end{eqnarray}
It is clear that teleportation is possible only for (18), (21) and not for (19) and (20). For successful teleportation Alice communicates the result to Bob through classical channel by one-cbit, so that he can make the required operations on his qubits, to get the desired state. Probability of success is half.
\subsubsection{Teleportation of two qubits unknown state through $|P_{1} \rangle $ state }
It is known that a general unknown two-qubit state cannot be teleportated using $|P_{1} \rangle$ as quantum resource. Probabilistic teleportation of two qubit non-maximally entangled state $ \alpha |00 \rangle +\beta [|01 \rangle+ |10 \rangle]$ is possible by using local operations at Alice's end ($|\alpha|^{2} + 2|\beta|^{2} = 1$).
We assume that the particles 3, 4  belong to Alice and the particles 5, 6 belong to Bob:
\begin{eqnarray}
|\Psi \rangle_{12} \otimes |P_{1} \rangle_{3456} =\frac {1} {2}  [\alpha |000001 \rangle_{123456} + \alpha |000010 \rangle_{123456} + \alpha |000100 \rangle_{123456} + \alpha |001000 \rangle_{123456} + \nonumber \\
\beta |010001 \rangle_{123456} + \beta |010010\rangle_{123456} +\beta |010100\rangle_{123456}+\beta |011000\rangle_{123456} \nonumber \\
+\beta |100001 \rangle_{123456} + \beta |100010\rangle_{123456} +\beta |100100\rangle_{123456}+\beta |101000\rangle_{123456}].
\end{eqnarray}
First, Alice applies a control-NOT with 1 as control qubit and 4 as the target qubit and then another control-NOT with 2 as control qubit and 4 as target qubit. Subsequently, she takes 3 as control qubit and 4 as target qubit for applying her last control-Not operation. Alice now performs a von Neumann measurement on particle 4 to obtain,
\begin{eqnarray}
\langle 1|_4 P_{1} \prime\rangle_{123456} =\frac {1} {\sqrt {2}} [ \alpha |00000 \rangle_{12356} + \alpha |00100 \rangle_{12356} + \beta |01001 \rangle_{12356} + \beta |01010\rangle_{12356} +\nonumber\\ \beta |10001 \rangle_{12356} + \beta |10010\rangle_{12356} ],
\end{eqnarray}
or
\begin{eqnarray}
\langle0|_4 P_{1} \prime\rangle_{123456} =\frac {1} {\sqrt {2}}  [\alpha |00001 \rangle_{12356} + \alpha |00010 \rangle_{12356} +\beta |01000\rangle_{12356}+\beta |01100\rangle_{12356}+\nonumber\\ \beta |10000\rangle_{12356}+\beta |10100\rangle_{12356}].
\end{eqnarray}
Teleportation is possible if the result is $|1_{4}\rangle$ and fails if it is $|0_{4}\rangle$. Alice now applies a Hadamard operation on particle 2 and then measures her three particles in computational basis. Teleportation becomes successful, if the outcome of measurement is $|000 \rangle_{123}$ or $|010 \rangle_{123}$ and fails otherwise. In table 3, outcomes of the final measurement of Alice in computational basis, which leads to successful teleportation are shown.\\
\begin{center}
\begin{table}[h]
\caption{\label{tab3} The outcome of the measurement performed by Alice which leads to successful teleportation and the state obtained by Bob}
\begin{tabular}{|c|c|}
\hline
{\bf Outcome of Alice's Measurement }&{\bf State obtained}\\
\hline
$|000\rangle$&$\alpha(|00_{56}\rangle)+\beta(|01_{56}\rangle+|10_{56}\rangle)$\\
$|010\rangle$&$\alpha(|00_{56}\rangle)-\beta(|01_{56}\rangle+|10_{56}\rangle)$\\
\hline
\end{tabular}
\end{table}
\end{center}
She then sends the result to Bob so that he can make the required operations on his qubit to get the desired state.\\
When she fails to teleport the non-maximally entangled unknown two qubit state, she can regain this state with certain probability by measuring particle 3 in computational basis. If outcome is $|0_{3}\rangle$ she classically communicates to Bob, Bob then applies a Hadamard on particle 6, followed by a measurement of his two particles in computational basis (as given in table 4). Alice can regain the desired unknown state only if the outcome of Bob's measurement is $|00_{56}\rangle$ or $|01_{56}\rangle$ and fails for $|11_{56}\rangle$ or $|10_{56}\rangle$. Bob will send the information about his measurement through one c-bit. Here probability of success is 1/4. \\
\begin{center}
\begin{table}[h]
\caption{\label{tab4} The outcome of the measurement performed by Bob and the state obtained by Alice}
\begin{tabular}{|c|c|}
\hline
{\bf Outcome of Bob's Measurement}&{\bf State obtained}\\
\hline
$|00\rangle$&$\alpha(|00_{12}\rangle)+\beta(|01_{12}\rangle+|10_{12}\rangle)$\\
$|01\rangle$&$\alpha(-|00_{12}\rangle)+\beta(|01_{12}\rangle+|10_{12}\rangle)$\\
$|00\rangle$&$(|00_{12}\rangle)$\\
$|00\rangle$&$(|00_{12}\rangle)$\\
\hline
\end{tabular}
\end{table}
\end{center}
\subsection{Teleportation Through $|P_{2} \rangle$ state}
\subsubsection{Teleportation of single qubit unknown state through $|P_{2} \rangle $ state }
Now we consider  $|P_{2} \rangle $ state as the entangled channel,
\begin{equation}
|P_{2} \rangle =  \frac {1} {\sqrt {5}} [ | 0000 \rangle_{3456}  + | 1111 \rangle_{3456}  + |0011 \rangle_{3456} + |0101 \rangle_{3456} + |0110 \rangle_{3456}].
\end{equation} 
Here the particles 3,4 and 5 belong to Alice and the particle 6 belongs to Bob. With the addition of unknown state of 1 qubit at the Alice's end, the quantum channel takes the form,
\begin{eqnarray}
|\Psi \rangle_1 \otimes |P_{2} \rangle_{3456} =\frac {1} {\sqrt {5}}  [\alpha |00000 \rangle_{13456} + \alpha |01111 \rangle_{13456}+ \alpha |00011 \rangle_{13456} + \alpha |00101 \rangle_{13456} +\nonumber\\\alpha |00110 \rangle_{13456}+ \beta |10000 \rangle_{13456} + \beta |11111\rangle_{13456} +\beta |10011\rangle_{13456}+\nonumber\\\beta |10101\rangle_{13456}+\beta |10110\rangle_{13456}]. 
\end{eqnarray} 
 Alice applies a control-NOT with 4 as target qubit and 1 as control qubit and then applies a Hadamard on 1. Subsequently, Alice measures qubit 3 through von Neumann measurement, resulting in
\begin{eqnarray}
\langle0_{3}|P_{2} \prime\rangle_{13456} =\frac {1} {2\sqrt {2}}  [\alpha |0000 \rangle_{1456} + \alpha |1000 \rangle_{1456} +\alpha |0011 \rangle_{1456}+\alpha |1011 \rangle_{1456} + \alpha |0101 \rangle_{1456}\nonumber\\+ \alpha |1101 \rangle_{1456} + \alpha |0110 \rangle_{1456} +\alpha |1110 \rangle_{1456}+ \beta |0010 \rangle_{1456} - \beta |1010\rangle_{1456}+\beta |0001\rangle_{1456}-\beta |1001 \rangle_{1456} \nonumber \\+ \beta |0111\rangle_{1456} -\beta |1111\rangle_{1456}+\beta |0100\rangle_{1456}-\beta |1100\rangle_{1456}], 
\end{eqnarray}
or
\begin{eqnarray} 
\langle1_3|P_{2} \prime\rangle_{13456} =\frac {1} {\sqrt {2}} [\alpha |0111 \rangle_{1456} + \alpha |1111 \rangle_{1456} + \beta |0101\rangle_{1456}-\beta |1101\rangle_{1456}]. 
\end{eqnarray} 
Teleportation is possible if the result is $|0_{3}\rangle$ and fails if it is $|1_{3}\rangle$. Alice performs a measurement of qubit 4 in computational basis, followed by the measurement of  qubit 1 and 5 together in computational basis ($|00_{15}\rangle$, $|10_{15}\rangle$, $|01_{15}\rangle$ and $|11_{15}\rangle$ ) and sends the information to Bob about her measurement by 2 cbit. Bob performs unitary transformation to get the desired state.
\subsubsection{Teleportation of two qubit unknown state through $|P_{2} \rangle $ state }
For two qubit teleportation through $|P_{2} \rangle $ state, we consider the general unknown state,
\begin{equation}
|\Psi \rangle_{12} = \alpha |00 \rangle_{12} + \beta |11 \rangle_{12} +\gamma |01 \rangle_{12}+\delta |10\rangle,
\end{equation}
where $|\alpha|^{2} + |\beta|^{2} + |\gamma|^{2} +|\delta|^{2} = 1 $
Probabilistic teleportation is possible by applying local operations only if $\alpha$=$\beta$ and $\gamma=\delta$ and it is worth mentioning that through standard protocol even if $\alpha$=$\beta$ and $\gamma=\delta$ teleportation may not be possible. The particles 3, 4 belong to Alice and the particles 5, 6 belong to Bob. Alice wants to teleport the state $\alpha[|00\rangle +|11\rangle]+\gamma[|01\rangle+|10\rangle]$ to Bob:
\begin{eqnarray}
|\Psi \rangle_{12} \otimes |P4 \rangle_{3456} =\frac {1} {\sqrt {5}}  [\alpha |000000 \rangle_{123456} + \alpha |000011 \rangle_{123456} + \alpha |001111 \rangle_{123456} + \alpha |000101\rangle_{123456}\nonumber\\+\alpha |000110 \rangle_{123456} + \alpha |110000 \rangle_{123456} + \alpha |110011 \rangle_{123456} + \alpha |111111 \rangle_{123456}  +\alpha |110101 \rangle_{123456} + \nonumber\\\alpha |110110 \rangle_{123456} + \gamma |010000 \rangle_{123456} + \gamma |010011\rangle_{123456} +\gamma |01111\rangle_{123456}+\gamma |010101\rangle_{123456} + \nonumber\\\gamma |010110 \rangle_{123456} + \gamma |100000\rangle_{123456} +\gamma |100011\rangle_{123456}+\gamma |101111\rangle_{123456} +\gamma |100101 \rangle_{123456} + \gamma |100110\rangle_{123456}].  
\end{eqnarray}
Alice applies control-NOT with 4 as target qubit and 1 as control qubit and control-NOT with 4 as target and 2 as control qubit. Subsequently, Alice performs von Neumann measurement of qubit 4, followed by another von Neumann measurement of qubit 3:
\begin{eqnarray}
\langle0_3\langle0_4|P_{2} \prime\rangle_{123456} =\frac {1} {\sqrt {2}}  [\alpha |0000 \rangle_{1256} + \alpha |0011 \rangle_{1256}+ \alpha |1100 \rangle_{1256} + \alpha |1111 \rangle_{1256}  + \nonumber \\+\gamma |0101\rangle_{1256}+\gamma |0110 \rangle_{1256} \nonumber \\+\gamma |1001 \rangle_{1256} + \gamma |1010\rangle_{1256}],  
\end{eqnarray} 
\begin{eqnarray}
\langle1_3\langle1_4|P_{2} \prime\rangle_{123456} =\frac {1} {{\sqrt {2}}\alpha}  [\alpha |0011 \rangle_{1256} + \alpha |1111 \rangle_{1256}],  
\end{eqnarray} 
\begin{eqnarray}
\langle1_3\langle0_4|P_{2} \prime\rangle_{123456} =\frac {1} {{\sqrt {2}}\gamma}  [\gamma |0111\rangle_{1256}+\gamma |1011\rangle_{1256}],  
\end{eqnarray}
or
\begin{eqnarray}
\langle0_3\langle1_4|P_{2} \prime\rangle_{123456} =\frac {1} {\sqrt {3|\alpha|^2 + 4|\gamma|^2}}  [\alpha |0001\rangle_{1256}+\alpha |0010 \rangle_{1256} + \alpha |1110 \rangle_{1256} + \gamma |0100 \rangle_{1256} + \gamma |0111\rangle_{1256} + \nonumber \\ + \gamma |1000\rangle_{1256} +\gamma |1011\rangle_{1256}].  
\end{eqnarray} 
Teleportation is possible with (31), (34) and not with (32), (33). For teleportation through (34), Alice applies Hadamard on particle 2 and measures both the particles in computational basis. Teleportation is successful for the outcome $|00\rangle$ and $|01\rangle$ and fails for $|11\rangle$ and $|10\rangle$. If Alice gets (31), then she needs to measure (31) in the basis given below,
\begin{eqnarray}
|\Phi_{1} \rangle = \frac{1}{2}[|+\rangle|+ \rangle ],  
\end{eqnarray}
\begin{eqnarray}
|\Phi_{2} \rangle = \frac{1}{2}[|-\rangle|+ \rangle ],  
\end{eqnarray} 
\begin{eqnarray}
|\Phi_{3} \rangle = \frac{1}{2}[|-\rangle|- \rangle ],  
\end{eqnarray}
and
\begin{eqnarray}
|\Phi_{4} \rangle = \frac{1}{2}[|+\rangle|- \rangle ].  
\end{eqnarray}\\
Alice then sends the outcome of the measurement to Bob as given in table 5, so that he can carry out the required unitary operations to get $\alpha(|00\rangle +|11\rangle) + \gamma(|01\rangle+ |10\rangle$).\\
\begin{table}[h]
\caption{\label{tab5} The outcome of the measurement performed by Alice and the state obtained by Bob}
\begin{center}
\begin{tabular}{|c|c|}
\hline
{\bf Outcome of Alice's Measurement}&{\bf State obtained}\\
\hline
$|\Phi_{1} \rangle = \frac{1}{2}[|+\rangle|+ \rangle ]$&$\alpha(|00_{56}\rangle + |11_{56}\rangle) + \gamma(|01_{56}\rangle + |10_{56}\rangle)$\\
$|\Phi_{2} \rangle = \frac{1}{2}[|-\rangle|+ \rangle ]$&$\alpha(|00_{56}\rangle - |11_{56}\rangle) + \gamma(|01_{56}\rangle - |10_{56}\rangle)$\\
$|\Phi_{1} \rangle = \frac{1}{2}[|-\rangle|- \rangle ]$&$\alpha(|00_{56}\rangle + |11_{56}\rangle) - \gamma(|01_{56}\rangle + |10_{56}\rangle)$\\
$|\Phi_{2} \rangle = \frac{1}{2}[|+\rangle|- \rangle ]$&$\alpha(|00_{56}\rangle - |11_{56}\rangle) + \gamma(-|01_{56}\rangle + |10_{56}\rangle)$\\
\hline
\end{tabular}
\end{center}
\end{table}
\subsection{Teleportation through $|P_{3} \rangle $ and  $|P_{4} \rangle $ states }
\subsubsection{Teleportation of single qubit through $|P_{3} \rangle $ state }
We now consider $|P_{3} \rangle $ as the quantum channel,
\begin{equation}
|P_{3} \rangle_{3456} =  \frac {1} {2} [ | 0000 \rangle_{3456}  + | 0101 \rangle_{3456}  + |1000 \rangle_{3456} + |1110 \rangle_{3456}].
\end{equation}
The particles 3,5 and 6 belong to Alice and the particle 4 belongs to Bob. With the addition of the unknown state of one qubit at Alice's end, the quantum channel takes the form, 
\begin{eqnarray}
|\Psi \rangle_1 \otimes |P_{3} \rangle_{3456} =\frac {1} {2}  [\alpha |0000 \rangle_{1356}|0\rangle_{4} + \alpha |0001 \rangle_{1356}|1\rangle_{4} + \alpha |0100 \rangle_{1356}|0\rangle_{4} + \alpha |0011 \rangle_{1356}|1\rangle_{4} + \beta |1000 \rangle_{1356}|0\rangle_{4} \nonumber \\
  + \beta |1001\rangle_{1356}|1\rangle_{4} +\beta |1100\rangle_{1356}|0\rangle_{4}+\beta |1011\rangle_{1356}|1\rangle_{4}]. 
\end{eqnarray} 
Above state leads to faithful teleportation by the standard protocol. It will be seen that by applying local operations teleportation becomes probabilistic. In the present case, Alice applies control-NOT with 1 as control qubit and 3 as target qubit and another control-NOT, with 1 as control and 4 as target qubit. Then Alice applies Hadamard on particle 1:
\begin{eqnarray}
|P_{3} \prime\rangle_{13456} =\frac {1} {2\sqrt{2}} [(|0000\rangle_{1345})(\alpha |0\rangle_{6}+\beta|1\rangle_{6})+(|1000\rangle_{1345})(\alpha |0\rangle_{6}-\beta|1\rangle_{6})+(|0011\rangle_{1345})(\alpha |1\rangle_{6}+\beta|0\rangle_{6})+\nonumber\\ (|1011\rangle_{1345})(\alpha |1\rangle_{6}-\beta|0\rangle_{6})+\alpha|0010\rangle_{1345}|1\rangle_{6} + \alpha|1010\rangle_{1345}|1\rangle_{6} +\alpha|0100\rangle_{1345}|0\rangle_{6} + \alpha|1100\rangle_{1345}|0\rangle_{6}    \nonumber\\+ \beta|0010\rangle_{1345}|1\rangle_{6}-\beta|1010\rangle_{1345}|1\rangle_{6}+ \beta|0111\rangle_{1345}|0\rangle_{6} -\beta|1111\rangle_{1345}|0\rangle_{6}].
\end{eqnarray} 
After that Alice performs a measurement in computational basis on her qubits and conveys the result to Bob so that he can make the required operations on his qubits to get the desired result. In this case probability of success is one-third. As is evident states $|0000\rangle_{1345}$, $|1000\rangle_{1345}$, $|0011\rangle_{1345}$ or $|1011\rangle_{1345}$ results in successful teleportation, while remaining fails.
\subsubsection{Teleportation of single qubit through $|P_{4} \rangle $ state }
Now we consider $|P_{4} \rangle $ as the entangled channel,
\begin{equation}
|P_{4} \rangle_{3456} =  [\frac {1} {2} ][ | 0000 \rangle_{3456}  + | 1011 \rangle_{3456}  + |1101 \rangle_{3456} + |1110 \rangle_{3456},
\end{equation}
where the particles 3,4 and 5 belongs to Alice and the particle 6 belongs to Bob. With the addition of unknown state of one qubit at the Alice's end, the quantum channel takes the form, 
\begin{eqnarray}
|\Psi \rangle_1 \otimes |P_{4} \rangle_{13456} =\frac {1} {2}  [\alpha |00000 \rangle_{13456} + \alpha |01011 \rangle_{13456} + \alpha |01101 \rangle_{13456} + \alpha |01110 \rangle_{13456} + \beta |10000 \rangle_{13456} \nonumber \\+ \beta |11011\rangle_{13456} +\beta |11101\rangle_{13456}+\beta |11110\rangle_{13456}]. 
\end{eqnarray}
Like the previous state, the above state leads to faithful teleportation by standard protocol and through local operations teleportation becomes probabilistic. Alice applies control-NOT with 1 as control qubit and 5 as target qubit. Subsequently, Alice applies Hadamard on qubit 1 and measures particle 1 in computational basis. Teleportation is possible if the outcome is $|1_{1}\rangle$ and fails if it is $|0_{1}\rangle$. The state, if the measurement outcome is $|1_{1}\rangle$, is given by,
\begin{eqnarray}
|P_{4} \prime\prime\rangle_{3456} =\frac {1} {2\sqrt{2}}[(|1110\rangle)(\alpha |1\rangle-\beta|0\rangle)+(|1111\rangle)(\alpha |0\rangle-\beta|1\rangle)+(|0111\rangle)(\alpha |0\rangle+\beta|1\rangle)\nonumber \\+(|0110\rangle)(\alpha |1\rangle+\beta|0\rangle)+\alpha|00000\rangle + \alpha|10000\rangle +\alpha|01011\rangle +\nonumber \\\alpha|11101\rangle +\beta|00010\rangle-\beta|10010\rangle+\beta|01011\rangle -\beta|11011\rangle].
\end{eqnarray}
Alice now performs a measurement in computational basis on her qubits and passes the results to Bob, so that he can make the required operations on his qubits to get the desired state. In this way, we can see that local operation leads to probabilistic result. For a general two qubit state, teleportation is not possible by standard protocol and also through the present scheme. 
\section{Conclusion}
In conclusion, we have demonstrated the usefulness of tripartite W-state and different quadripartite entangled states as quantum resource for teleportation using an unconventional teleportation scheme. We make use of entanglement monogamy\cite{28} to incorporate an unknown state in a multipartite entangled channel, such that the receiver partially gets disentangled from the network. We hope this procedure finds experimental verification. W-state and a number of four particle entangled states earlier found unsuitable for teleportation, can teleport probabilistically through this scheme. It is also found out that for certain four particle channels, the present procedure does not succeed, although the conventional approach works well.
\section*{Acknowledgment}
The authors acknowledge S. Muralidharan for many useful discussions.


\begin{thebibliography}{28}
\bibitem{1}C. Bennett, G. Brassard, C. Crepeau, R. Jozsa, A. Peres, and W. K. Wooters, Phys. Rev. Lett. \textbf{70}, 1895 (1993).

\bibitem{2}C. H. Bennett and S. Wiesner, Phys. Rev. Lett. \textbf{69}, 2881 (1992).

\bibitem{3}M. Zukowski, A. Zeilinger, M. A. Horne and A. K. Ekert, Phys. Rev. Lett. \textbf{71}, 4278 (1993).

\bibitem{4}M. Hillery, V. Buzek, and A. Berthiaume, Phys. Rev. A \textbf{59}, 1829 (1999).

\bibitem{5}N. Gisin, G. Ribordy, W. Tittel, and H. Zbinden, Rev. Mod. Phys. \textbf{74}, 145 (2002).

\bibitem{6} S. Jain, S. Muralidharan and P. K. Panigrahi, EPL \textbf{87}, 60008 (2009).

\bibitem{7}J. R. Samal, M. Gupta, P. K. Panigrahi and A. Kumar, J. Phys. B: At. Mol. Opt. Phys. \textbf{43}, 095508 (2010).

\bibitem{8}M. A. Nielsen and I. L. Chuang, Quantum Computation and Quantum Information (Cambridge Univ. Press, 2002).

\bibitem{9}D. Bouwmeester, J. W. Pan, K. Mattle, M. Eibl, H. Weinfurter, and A. Zeilinger Nature \textbf{390}, 575 (1997).

\bibitem{10}Q. Zhang, A. Goebel, C. Wagenknecht, Y. A. Chen, B. Zhao, T. Yang, A. Mair, J. Schmiedmayer and J. W. Pan, Nature \textbf{2}, 678 (2006).

\bibitem{11} D. Boschi, S. Branca, F. De Martini, L. Hardy and S. Popescu, Phys. Rev. Lett. \textbf{80}, 1121 (1998).

\bibitem{12}S. Muralidharan and P. K. Panigrahi, Phys. Rev. A \textbf{77}, 032321 (2008).

\bibitem{13}S. Prasath, S. Muralidharan, C. Mitra and P. K. Panigrahi, QINP, DOI 10.1007/s11128-011-0252-z (2011).

\bibitem{14}S. Choudhury, S. Muralidharan and P. K. Panigrahi, J. Phys. A: Math. Theor. \textbf{42}, 115303 (2009).

\bibitem{15}S. Muralidharan, S. Karumanchi, S. Jain, R. Srikanth, P. K. Panigrahi, Euro. Phys. J. D \textbf{61}, 757 (2011).

\bibitem{16}S. Muralidharan and P. K. Panigrahi, Phys. Rev. A \textbf{78}, 062333 (2008) and reference therein.

\bibitem{17}A. K. Pati and P. Agrawal, Phys. Lett. A \textbf{371}, 185 (2007).

\bibitem{18}V. N. Gorbachev, A. I. Trubilko and A. A. Rodichkina, Phys. Lett. A \textbf{314}, 267 (2003).

\bibitem{19}W. Dur, G. Vidal, and J. I. Cirac, Phys. Rev. A \textbf{62}, 062314 (2002).

\bibitem{20}F. Verstreate, J. Dehaene, B. De Moor and H. Verschelde, Phys. Rev. A \textbf{65}, 052112 (2002).

\bibitem{21}D. Li, X. Li, H. Huang and X. Li, Phys. Rev A \textbf{76}, 052311 (2009). 

\bibitem{26}B. Pradhan, P. Agrawal and A. K. Pati, quant-ph/0705.191 (2008).

\bibitem{27}L. Lamata, J. Leon, D. Salgado and E. Solano, Phys. Rev. A \textbf{75}, 022318 (2007).

\bibitem{22}B. S. Shi and A. Tomit, Phys. Lett. A \textbf{296}, 161 (2002).

\bibitem{23}J. Joo, Y. Park, S. Oh and J. Kim, New J.Phys. \textbf{5}, 136 (2003).

\bibitem{24}Z. L. Cao and W. Song, Physica A: Statistical Mechanics and its Applications, \textbf{347}, 177 (2005).

\bibitem{25}S. Muralidharan, S. Karumanchi, S. Narayanaswamy, R. Srikanth and P. K. Panigrahi, quant-ph/0907.3532v2.(2011).

\bibitem{28}V. Coffman, J. Kundu, and W. K. Wootters, Phys. Rev. A \textbf{61}, 052306 (2000).
\end{thebibliography}
\end{document}